\documentclass[a4paper,12pt]{article}
\usepackage{longtable}
\usepackage{epsf,amsmath}
\usepackage{indentfirst}
\usepackage{amssymb}
\usepackage{hhline}
\usepackage{array,tabularx}
\def\@biblabel#1{#1.}
\makeatother

\begin{document}

\title{\large\bf An Exo-Jupiter Candidate in the Eclipsing Binary FL Lyr}

\author{V. S. Kozyreva$^1$, A. I. Bogomazov$^{1}$, B. P. Demkov$^{2,3}$, \\ L. V. Zotov$^1$, A. V. Tutukov$^4$ \\
\small\it $^1$Sternberg Astronomical Institute, Lomonosov Moscow State University,\\
\small\it Universitetskii pr. 13, Moscow, 119991 Russia \\
\small\it $^2$``IT Project'', Savelkinskii proezd 4, Zelenograd, Moscow, 124482 Russia \\
\small\it $^3$All-Russian Research Institute of Physical, Technical,\\
\small\it and Radio Technical Measurements, \\
\small\it Mendeleevo, Moscow Region, 141570 Russia \\
\small\it $^4$Institute of Astronomy, Russian Academy of Sciences, \\
\small\it ul. Pyatnitskaya 48, Moscow, 119017 Russia \\ }

\date{\begin{minipage}{15.5cm} \small
Light curves of the eclipsing binary FL Lyr acquired by the Kepler space telescope are analyzed.
Eclipse timing measurements for FL Lyr testify to the presence of a third body in the system. Preliminary
estimates of its mass and orbital period are $\gtrsim 4M_J$ and $\gtrsim 7$ yrs. The times of primary minimum in the light
curve of FL Lyr during the operation of the Kepler mission are presented.
\end{minipage}}

\maketitle

\section{INTRODUCTION}

Extra-solar planetary systems remained hypothetical objects until the 1990s, when modern means for their detection were developed. Since then, some $10^4$ candidate exoplanets have been discovered using various methods; the existence of many of these exoplanets has been reliably confirmed \cite{exocatalogue}. The vast majority of the discovered planets orbit single stars or individual components of wide multiple systems.

Currently, we know of eight exoplanets in five stellar systems and two candidate planets that simultaneously orbit both components of binaries, with both stars on the main sequence. The first planet discovered in such a binary was Kepler-16 (AB)b \cite{doyle2011}. Others include Kepler-34 (AB)b and Kepler-35 (AB)b \cite{welsh2012}, Kepler-38 (AB)b \cite{orosz2012a}, Kepler-47 (AB)b è Kepler-47 (AB)c \cite{orosz2012b}, PH1-Kepler-64 b \cite{schwamb2013}, Kepler-413 (AB)b \cite{kostov2014a,kostov2014b}, a possible third planet in the Kepler-47 \cite{hinse} system and the candidate planet KIC 9632895 (AB)b \cite{welsh-10}. Several planets near cataclysmic variables and a planet near the young star FW Tau \cite{exocatalogue} have also been discovered.

Searching for planets in binary systems is important for a number of reasons. Though it follows from \cite{holman1999} that planetary orbits in binaries exhibit long-term stability, it remains to be confirmed from observations that planets can survive in systems with various parameters. The systems known
up to now have very similar parameters. The presence or absence of planets in binary systems and
the systems` parameters are very important for our understanding of the processes of star and planet
formation (e.g., \cite{tutukov2012a}). In addition, binary systems are more favorable for harboring life than single stars, and could in principle have several inhabited planets \cite{mason2013}. This makes searches for planets in various binary systems very important for searches for extraterrestrial,
possibly even intelligent, life. A list of binary stars suitable for planetary searches can be found in \cite{tutukov2012b}, and includes the FL Lyr system.

In 2009-2014, the area of the sky containing FL Lyr was in the field of view of the Kepler space telescope \cite{kepler-fov}, which was launched into near-solar orbit with the aim of searching for exoplanets. During these years, the telescope carried out a continuous photometric sky survey, during which a large amount of observing material was accumulated for FL Lyr. The Kepler observations can be used to study various scientific problems. In particular, a third body orbiting an eclipsing variable star gives rise to periodic shifts of the system's center of mass with respect to the observer, causing the observed orbital period of the binary to vary about a certain value. The aim of our current study is to study the light-time effect in the FL Lyr system\footnote{In other words, to perform timing of the minima of the FL Lyr
light curve.}.

\section{THE ECLIPSING BINARY FL Lyr}

The eclipsing variable FL Lyr was discovered on photographic plates in 1935 \cite{morgenroth1935}. Its minima are deep, with the change in the star's brightness at the primary and secondary minimum being different by a factor of two: $m_{max}=9^{m}.27$, $m_{min \; I}=9^{m}.89$, $m_{min \; II}=9^{m}.52$. According to the ``General Catalog of Variable Stars'' \cite{okpz}, $P_{orb}=2.^{d}1781544$. The stars
in the system have different sizes and spectral types. In the 1950s, Struve \cite{struve1950} obtained a spectroscopic radial-velocity curve of the primary and determined its spectral type to be G5.

In 1963, Cristaldi \cite{cristaldi1965} obtained a photoelectric light curve and derived the photometric parameters of the system. He was able to estimate the masses of both components using earlier spectroscopy from various studies, both published and unpublished. The component parameters he found suggest that the two stars form an Algol-type system. In such systems, the primary (initially the higher-mass component)
has left the main sequence (MS) and begun its expansion; in the process, the star has transferred some
of its mass to the secondary. The mass of the primary becomes lower and its radius larger than its companion. At the same time, the primary is still on the MS, and the primary's luminosity is lower than the secondary's. This star is thus erroneously taken to be the secondary, while its companion, which initially had lower mass and is on the MS, is taken to be the primary.

Cristaldi \cite{cristaldi1965} presents parameters calculated for the FL Lyr system: one of the components has a mass of $M_1= 1 M_{\odot}$, a relative radius\footnote{ Expressed in fractions of the orbital semi-major axis of FL Lyr. } $r_1=0.132$ and a relative luminosity\footnote{ Expressed in fractions of the system's combined luminosity. } in the V filter $L_1 = 0.234$, while the second component has $M_2 = 1.1 M_{\odot}$, $r_2 = 0.114$, $L_2 = 0.430$. This thus appears to be an Algol-type system, when the lower-mass star has a larger radius than the higher-mass one. A so-called ``third light'', $L_3 = 0.336$, is also present in the solution, and belongs to either a field star or a third component in the system. The binary components move in a plane almost orthogonal to the plane of the sky; the estimated product of the eccentricity and longitude of periastron $e\cdot cos\omega\le 0.0002$. The orbits of the components are essentially circular. According to \cite{cristaldi1965}, the spectral types of the stars are estimated to be G0V + G5V.

In this study of the FL Lyr light curve, Botsula \cite{botsula1978} found that the system's light decreased near the secondary minimum. She proposed that the system contained diffuse matter situated close to the secondary, which blocks some of the light in the secondary eclipse and distorts the shape of this part of the curve. This hypothesis is fully in accord with the physics of Algol-type stars, where the envelope of the more evolved primary flows toward the secondary. Moreover, the FL Lyr light curve displayed episodes of fading by as much as $0.^m04-0.^m05$. It is possible that all these distortions have a random character, due to systematic errors in the particular part of the FL Lyr light curve studied in
\cite{botsula1978}.

In 1986, Popper et al.\cite{popper1986} obtained photoelectric
light curves and spectroscopic radial-velocity curves
of FL Lyr. They derived a new photometric solution of the light curve
($r_1=0.140$, $r_2=0.105$, $i =86^{\circ}.3$, $L_1=0.79$, $L_2 = 0.21$), and estimated the stellar masses and spectral types: $1.22 M_{\odot}$, $0.96 M_{\odot}$, F8+G8. The binary orbit is circular with high accuracy. Comparing the observed parameters of the system to those determined from theoretical evolutionary tracks of stars of the same mass, they estimated the age of the FL Lyr system to be $5.3\cdot 10^7$--$3.55\cdot 10^9$ yrs, with the most likely age being $2.29\cdot 10^9$ yrs \cite{lastennet2002}. Since the MS lifetime of a star with a mass of $1.2M_{\odot}$
is approximately $4.9\cdot 10^9$ (Eq. 6 in \cite{lipunov2009}), neither component of FL Lyr has left the MS. No third light was detected in \cite{popper1986}. When calculating the photometric parameters, an upper limit $k\le 1$ was imposed on the parameter $k=r_2/r_1$ (the ratio of the radii of the secondary and the primary). This excludes all solutions for which the two stars form an Algol-type system, i.e., a system with a reversed component-radius ratio. However, Popper et al. \cite{popper1986} suggest that the correctness of their derived parameters is supported by the lack of systematic deviations between the calculated and observed values for the brightness difference as a function of time. The parameters of the stars differ considerably from the solution found by Cristaldi \cite{cristaldi1965}. Among the characteristic features of the light curves, Popper et al. \cite{popper1986} noted a brightness modulation ($\Delta m =0.^{m}007$), which they attributed to the axial rotation of the components.

\section{OBSERVATIONS OF FL Lyr WITH THE KEPLER SPACE TELESCOPE}

We studied data obtained with Kepler. The main goal of the Kepler project was to search for exoplanets
using observations of their transits. We used the Kepler data for eclipse-timing measurements (determining the light-time effect) for FL Lyr. Detailed information on the Kepler space telescope can be found in \cite{kepler-description-1}.

The Kepler data we used can be found in the Barbara A. Mikulski Archive for Space Telescopes \cite{kepler-archive}, which is supported by the Space Telescope Science Institute. The identification number of FL Lyr
in the Kepler Input Catalog is 9641031. Detailed information on the search and retrieval of data from
the archive can be found in \cite{kepler-description-1}.

The Kepler archive consists of data files in FITS format. Two versions of these files are provided: LC
(long cadence) and SC (short cadence). LC is the main version; these data were collected once each
30 minutes. One LC FITS file contains observations of one object over one quarter\footnote{The light curves within the FL Lyr minima contain only five to six data points in the LC mode.}. SC (short cadence) SC is a complementary version of the data (intended for variability and asteroseismology studies); these data were collected once each minute. A single SC FITS file provides data for one month for a single object. Because of the design of Kepler, SC data were not accumulated during every quarter of the telescope's operation. SC data for FL Lyr are fully available only for the observing quarters 7, 8, 13, 14, 15, and 16. To improve the accuracy of our study, we used the FITS files obtained in SC mode. We converted the FITS files to a form convenient for the analysis using the IRAF software with the PyRAF extension (the kepconvert routine, which converts FITS files to text files).

\section{THE LIGHT-TIME EFFECT IN FL Lyr}

One of the methods that can be used to detect a third body in an eclipsing system is to search for the
light-time effect. The periods of the primary and secondary minima will oscillate if the distance between
the center of the solar system and the center of the eclipsing system varies. Any third body in a binary
system makes the system's center of mass move with the period of this body's orbit\footnote{The stability of planetary orbits in binary systems was studied in \cite{holman1999}; according to Table 6 in \cite{holman1999}, the orbit of a planet around the central binary will be stable if the semi-major axis
of the planet's orbit is approximately a factor of four or more larger than the semi-major axis of the binary orbit; i.e., if the orbital period of the planet is longer than the orbital period of the central binary star by a factor of 10 or more, as follows from Kepler's third law. Thus, the conditions for the long-term survival of planets in the FL Lyr system are satisfied for planetary orbital periods exceeding 20 days.}. The comparatively short period of the stars' orbit about the common center (about two days) makes it possible to identify a large number of light curves within minima in the Kepler observations. Our aim is to look for the light-time effect in the FL Lyr system; i.e., to search for shifts in the observed times of minima relative to the calculated values.

The orbit of a binary system rotates due to tidal forces between the two stars and general relativistic
effects. In the case of an elliptical orbit, this rotation is manifest through apsidal motion, which gives rise to a shift of the observed relative to the calculated times of minima. For FL Lyr, with its practically circular binary orbit ($e\le 0.0002$), theoretical estimates of the apsidal motion predict the period of this motion to be longer than 100 years, and its amplitude to be below 10 s. This effect is very small over the time interval of the Kepler observations, can be neglected. We are looking for the light-time effect with much shorter periods.

We considered solely the primary minima. These are symmetric and deep ($\approx 0.^{m}6$) -- more than twice as deep as the secondary minima -- increasing the accuracy of the timing of the minima by the same
factor. We identified 600 Kepler light curves within primary minima of FL Lyr.

Kepler observations possess systematic errors (see, for instance, section 7.1 in \cite{kepler-description-1}) -- so-called linear trends, which can reach several hundredths of a
magnitude during the duration of a minimum in the FL Lyr light curve. Our study of the FL Lyr light
curves already corrected for this linear trend using correction factors shows that the trend was not fully
removed, so that the shape of the light curves during minima remains distorted. The brightness difference
between the ingress and egress reaches several ten-thousandths of a magnitude, with different signs
for different light curves. Distortions by one ten-thousandth of a magnitude for the primary minima
of FL Lyr translate into an error of approximately 2 s in the times of minima. This is a large error, and is able to completely distort the light-time effect due to the presence of a planet with an amplitude of the order of 5-6 s.

Light-time effects for 1279 close eclipsing binary
systems were studied by Conroy et al. \cite{conroy2014}, who report
the discovery of 236 eclipsing binaries with suspected
third bodies. This study will be supplemented with
an analysis of orbital-period variations (Orosz et al.,
in preparation; see \cite{conroy2014}). It will be very interesting to
compare our results to those of Orosz et al.

We studied individual light curves of primary minima of FL Lyr obtained with Kepler and calculated
individual correction factors to remove the linear trend in each case, based on the hypothesis that the brightnesses at the eclipse ingress and egress are intrinsically the same. Since changes in the trend correction factors occur only rarely, no more than twice during the period of FL Lyr, we adopted the hypothesis that the trend does not curve significantly within a light-curve minimum ($\approx 4$ hours), and corrected the light curve using a single set of correction coefficients that were specific to each minimum. Even if the origins of the different brightnesses at the beginnings and ends of minima are physical, our approach enables us to search for the light-time effect, since we are studying the dynamics of changes in the times of minima. If the light curve changes slowly compared to the observing period, all the times of minima will be displaced relative to the true times, but the magnitude of this shift will not influence the amplitude of the light-time effect. If, however, the changes in the light curve are comparable to the duration of the observations, which is very improbable, the light-time effect
will be determined with additional systematic errors. Nevertheless, the period of the light-time effect can be derived from the light curve. When determining the times of the primary minima, we used a template
theoretical light curve calculated from the orbital elements and the relative parameters of the stars in this binary system.

\begin{figure}
\hspace{0cm} \epsfxsize=0.99\textwidth\centering \epsfbox{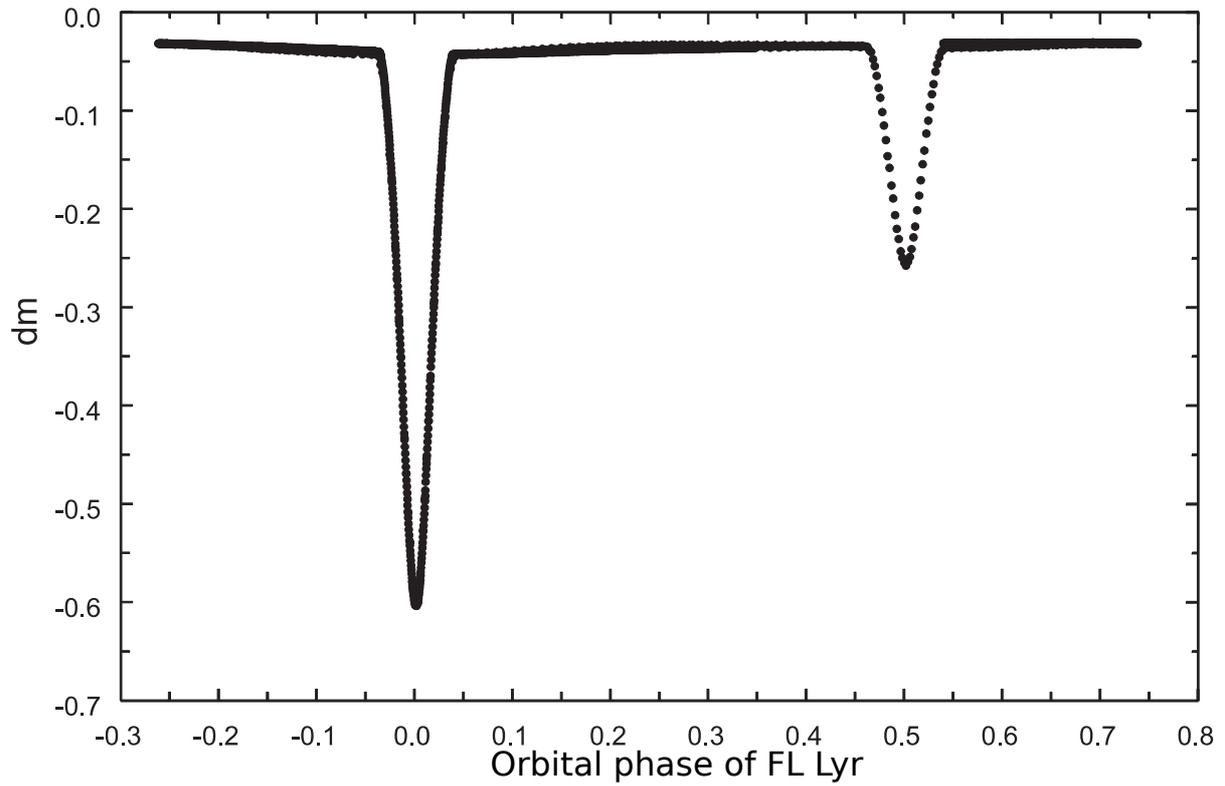}
\vspace{0cm}\caption{Light curve of FL Lyr compiled from Kepler observations between HJD 55031.54198 and HJD 55042.44777. It includes the times of primary minimum HJD 55032.54697, HJD 55034.72509, HJD 55036.90328, HJD 55039.08140, and HJD 55041.25957. All HJD times actually correspond to HJD–2400000.} \label{light}
\end{figure}

When calculating the photometric elements, we used the combined light curves compiled from observations obtained in the SC mode; these contained only neighboring observations within primary and secondary minima (Fig. \ref{light} shows a sample light curve of FL Lyr). The photoelectric light curves of FL Lyr obtained by Popper et al. \cite{popper1986} exhibit the same out-of-eclipse brightness level for both minima. We used this finding when selecting light curves used to calculate the photometric elements. Since the observations are distorted by the linear trend, we tried to select light curves for which this was minimal. The influence of the trend in each case is a random value, and the resulting sets of elements had nearly normal distributions. The parameter most important in the search for the light-time effect is the shift of the observed times of primary minimum relative to the calculated times (O-C). These shifts depended only weakly on variations of the other parameters we determined, presented in Table \ref{table1}. Column 2 of this table contains the ranges of the parameters found for the various light curves of FL Lyr. Column 3 gives the set of parameters we used to derive the theoretical curve we then applied as a template.

The CCD chip of the spacecraft has a broad filter ranging from 430 to 890 nm, corresponding to the
combined range of the B, V , and R filters of the Johnson photometric system. Therefore, it is not correct to compare photometric elements based on these light curves with elements derived using light curves obtained in the Johnson filters; this is especially true for the luminosities of the components.

We used a quasi-Newtonian method with analytical computation of the functional derivatives as
a minimization algorithm \footnote{The same algorithm was used earlier in \cite{khaliullin1983}--\cite{kozyreva2006}, resulting in the discoveries of brown dwarfs in the HP Aur and AS Cam systems.}. The minimization functional contains the sum of the squared differences
between the observed and theoretical magnitudes at each point, including simple and linear limitations
for the parameter values we seek. Because of their very weak influence on the light curves, we did not
vary the limb-darkening coefficients $u_1$ and $u_2$, and fixed them in accordance with the spectral types of the binary components (F8V + G8V \cite{popper1986}). Values for
the theoretical coefficients $u_1$  and $u_2$ corresponding to wavelengths in the middle of the instrumental range were taken from \cite{hamme1993}. Some of the parameters we determined in our free search for the orbital elements and parameters differ considerably from those obtained by Popper \cite{popper1986}; this is especially true for component luminosities. This can partially be explained
by the different spectral ranges used. In our current study, we are interested in the set of elements only
as a tool for deriving a theoretical curve that most closely approaches the observed curves at the primary minima.

\begin{figure}
\hspace{0cm} \epsfxsize=0.99\textwidth\centering \epsfbox{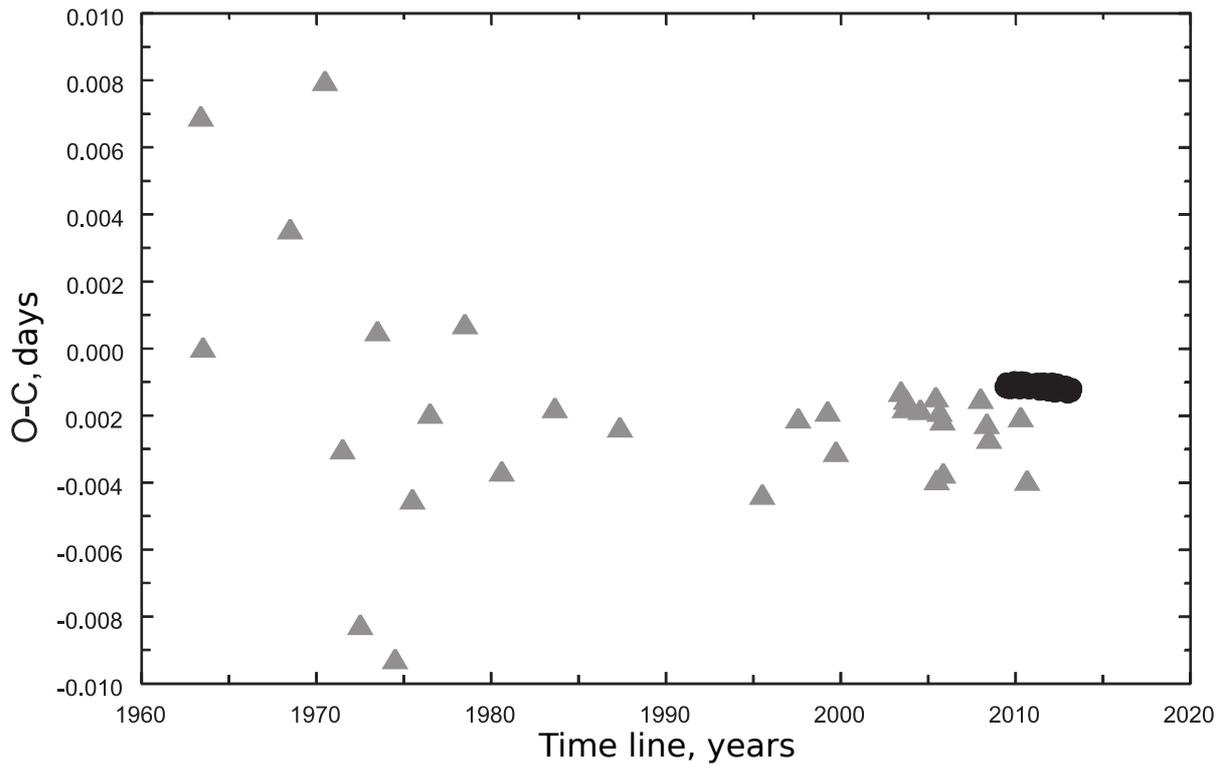}
\vspace{0cm}\caption{Times of minima of FL Lyr. The black circles show Kepler data (Table \ref{table3}), and the gray triangles data from ground-based
observations (Table \ref{table2} ). The x axis plots the dates of the observations and the y axis the differences between the observed times
of minimum and times of minimum calculated using \ref{eph-cris}.} \label{hist}
\end{figure}

Times of minima we collected from the literature are presented in Fig. \ref{hist} and Table \ref{table2}. The scatter of the data points in Fig. $\pm 15$ minutes. The scatter of the photoelectric times of minima can reach $\pm 1.5$ minutes. This large scatter of the times of minima can be explained in many ways, some of them described above. Our aim was to find the amplitude of the light-time effect with an accuracy of seconds. The large scatter in the previously published times of minima makes those data unsuitable for this. Accordingly, we used only the long uniform series of Kepler observations, deriving the times of minima using the same algorithm.

We used only data points corresponding to primary minima of FL Lyr (phases from 0.94 to 1.06). After
obtaining the light curve for an individual minimum without the linear trend, we then calculated the shift
of the observed time of minimum from the calculated time. We applied an algorithm for calculating the
minimal deviation between the observed and theoretical light curves\footnote{ See \cite{khaliullin1983}--\cite{kozyreva2006} for details.}; the only free parameter was the shift
of the primary minimum, with all other parameters being fixed at the values indicated in column 3 of Table \ref{table1}. The criterion for our solution was a symmetric position of the deviations (between the observed and calculated light-curve points) relative to zero phase. We checked this by determining the linear trend in the O-C residuals, with the result being considered satisfactory only in the absence of any trend. This procedure was performed for all primary minima of FL Lyr observed with Kepler; the times of minima are collected in the first column of \ref{table3}, while the second column contains the O-C residuals: the differences between the observed times of minima and the theoretical times of minima calculated with the ephemeris (\ref{eph-cris}).

Searching for the light-time effect requires as accurate as possible knowledge of the binary's orbital
period, on which the parameters of the light-time effect depend. We used three values of the orbital
period of FL Lyr. The period $P_{10} = 2.17815440^d$ was taken from \cite{cristaldi1965} (this is also the period given in \cite{okpz}). The period $P_{11} = 2^{d}.17815408^d$ data. Finally, the orbital period
$P_{13}=2^{d}.17815414$ was calculated using all the ground-based observations and six Kepler times of minima\footnote{So that the space data would not dominate the other measurements, we took three Kepler times of minima for 2009 and three for 2014. The 2009 times of minima are HJD-2400000 = 54965.02424, 54967.20240, and 54969.38054, and the 2014 times are HJD-2400000 = 56385.18082, 56387.35900, and 56389.53716.}. The system ephemerides for these three periods are

\begin{equation}
Min\; I (HJD) = 2438221.55250 + 2.17815440\times E;
\label{eph-cris}
\end{equation}

\begin{equation}
Min\; I (HJD) = 2438221.55239 + 2.17815408\times E; \label{ff3}
\end{equation}

\begin{equation}
Min\; I (HJD) = 2438221.55211 + 2.17815414\times E; \label{ff4}
\end{equation}

The gray triangles in Fig. 2 are the O-C values calculated with the ephemeris (\ref{eph-cris}) for the times of minima from the literature; the black circles are our values calculated using the Kepler observations. The shift in the times of minima we are seeking is clearly visible.

\begin{figure}
\hspace{0cm} \epsfxsize=0.99\textwidth\centering \epsfbox{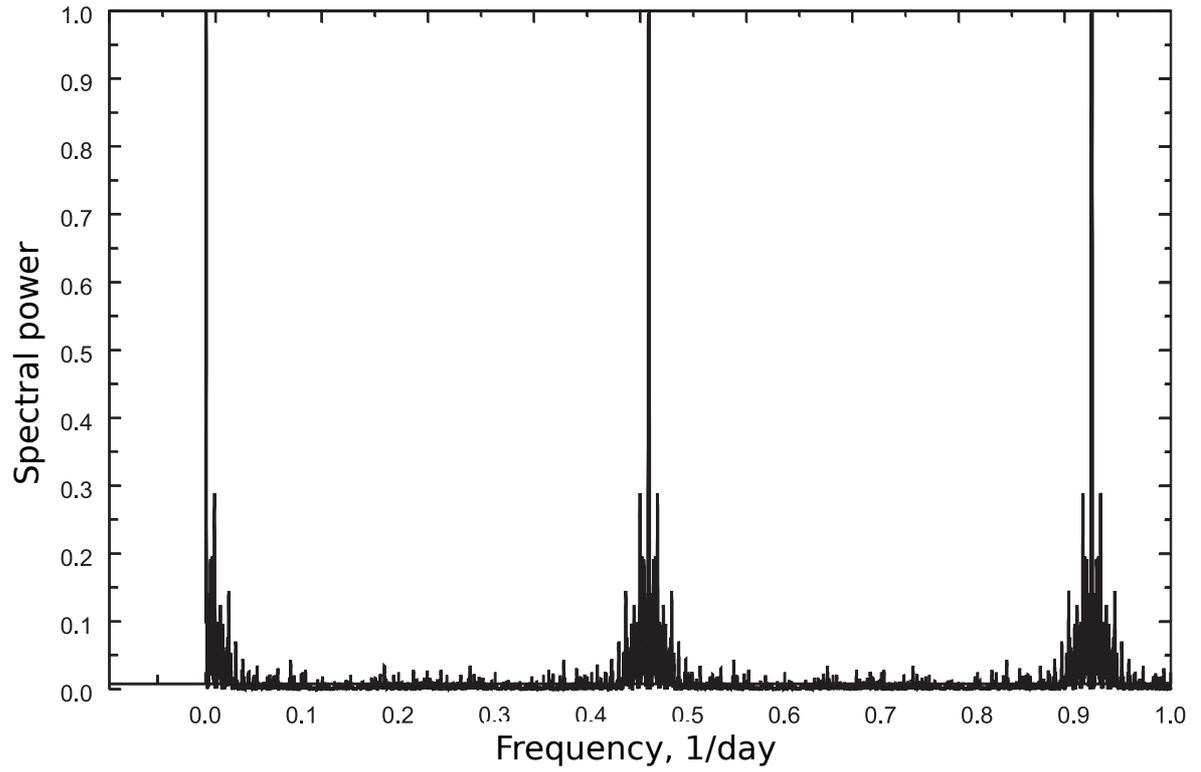}
\vspace{0cm}\caption{ Power spectrum of FL Lyr in relative units calculated using the ephemeris \ref{eph-cris}. } \label{pow1}
\end{figure}

\begin{figure}
\hspace{0cm} \epsfxsize=0.99\textwidth\centering \epsfbox{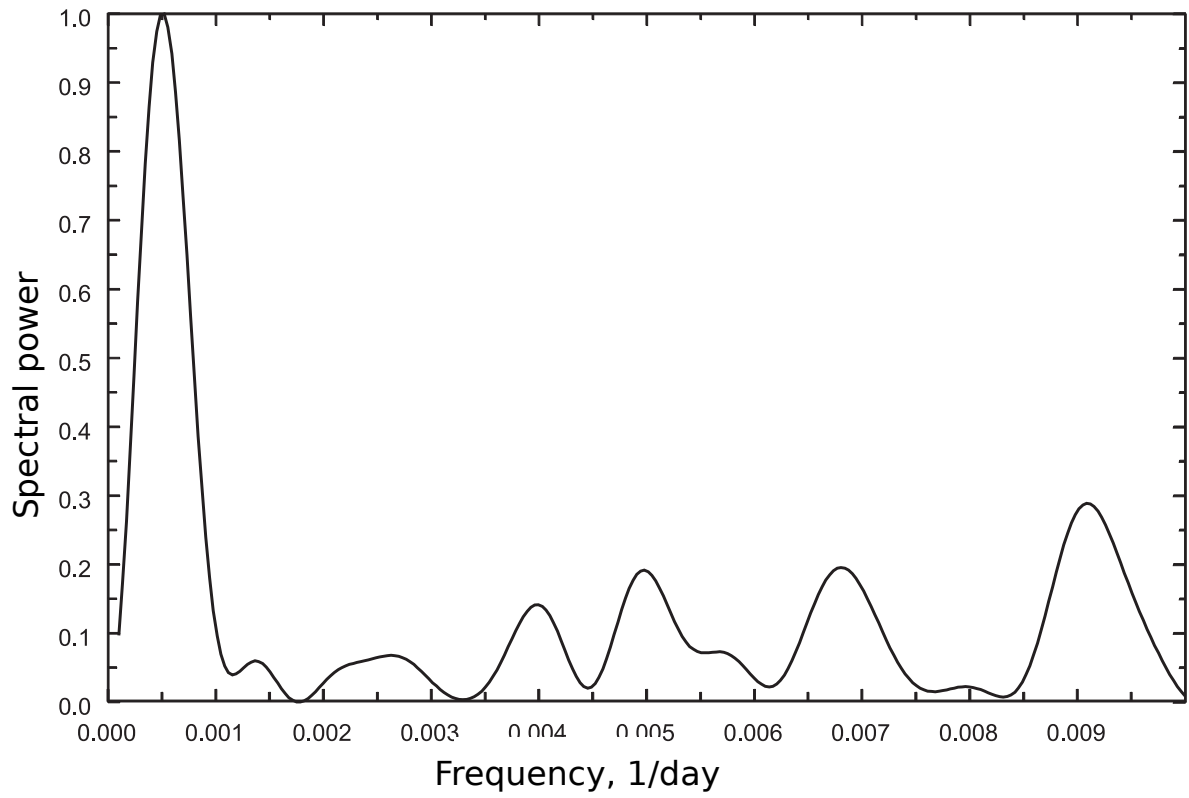}
\vspace{0cm}\caption{ Same as Fig. \ref{pow1} on a larger scale, for the part at low frequencies.. } \label{pow2}
\end{figure}

\begin{figure}
\hspace{0cm} \epsfxsize=0.99\textwidth\centering \epsfbox{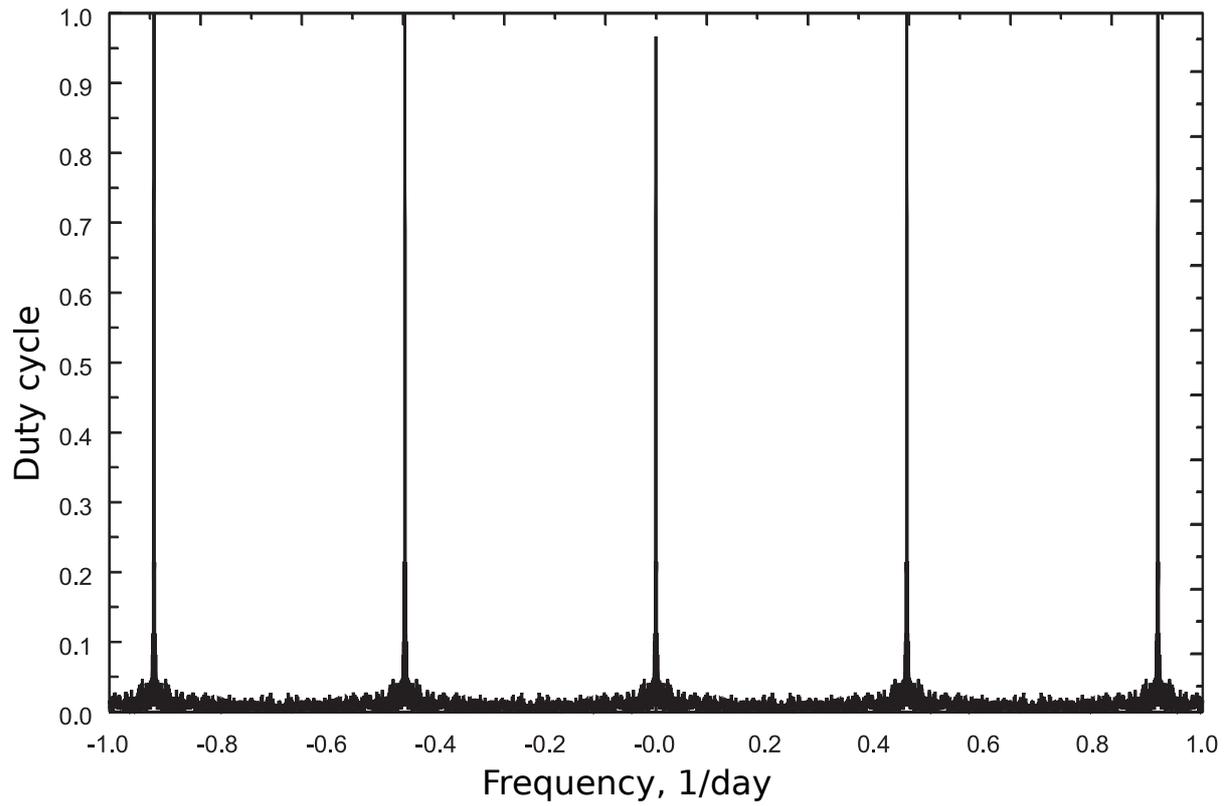}
\vspace{0cm}\caption{ Duty cycle as a function of the signal frequency.
} \label{skvazhnost}
\end{figure}

\begin{figure}
\hspace{0cm} \epsfxsize=0.99\textwidth\centering \epsfbox{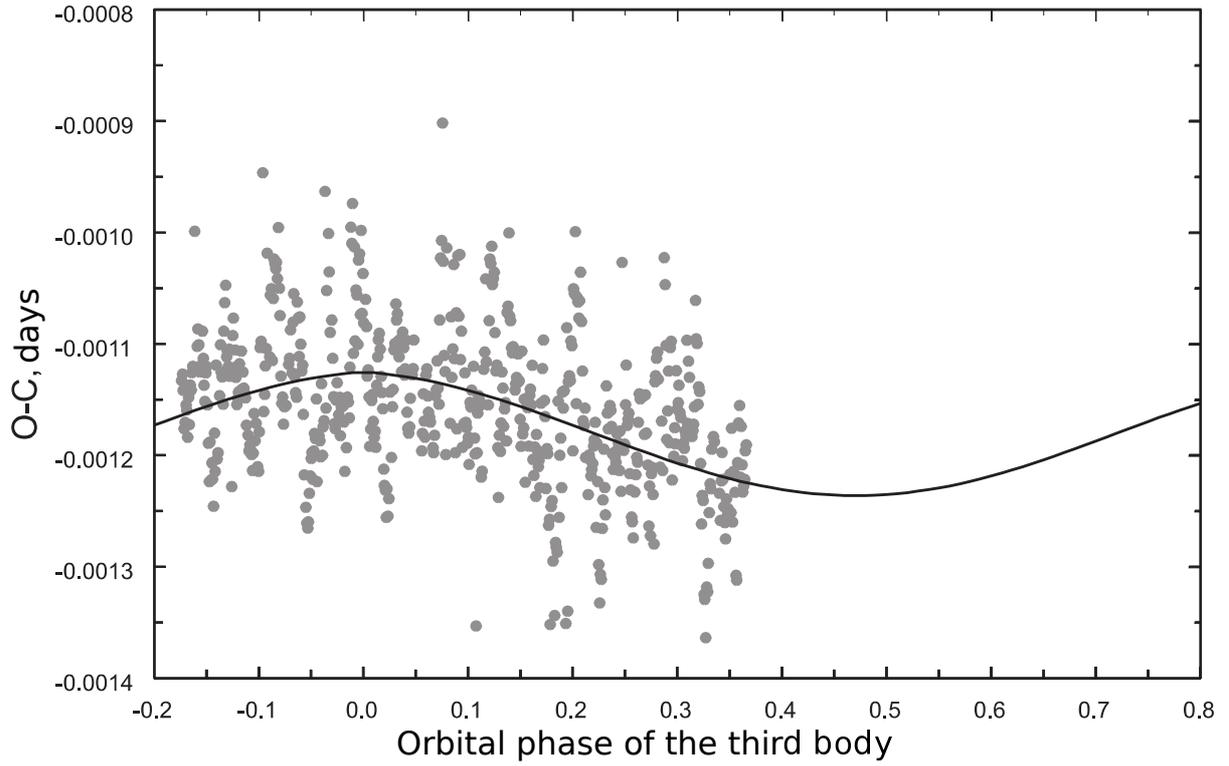}
\vspace{0cm}\caption{Light-time effect for the third body. The ephemeris \ref{eph-cris} was used in the calculations. See Table \ref{table4} for the period and amplitude of the light-time effect. The solid curve shows the theoretical curve, and the gray circles the observations.}
\label{p1}
\end{figure}

\begin{figure}
\hspace{0cm} \epsfxsize=0.99\textwidth\centering \epsfbox{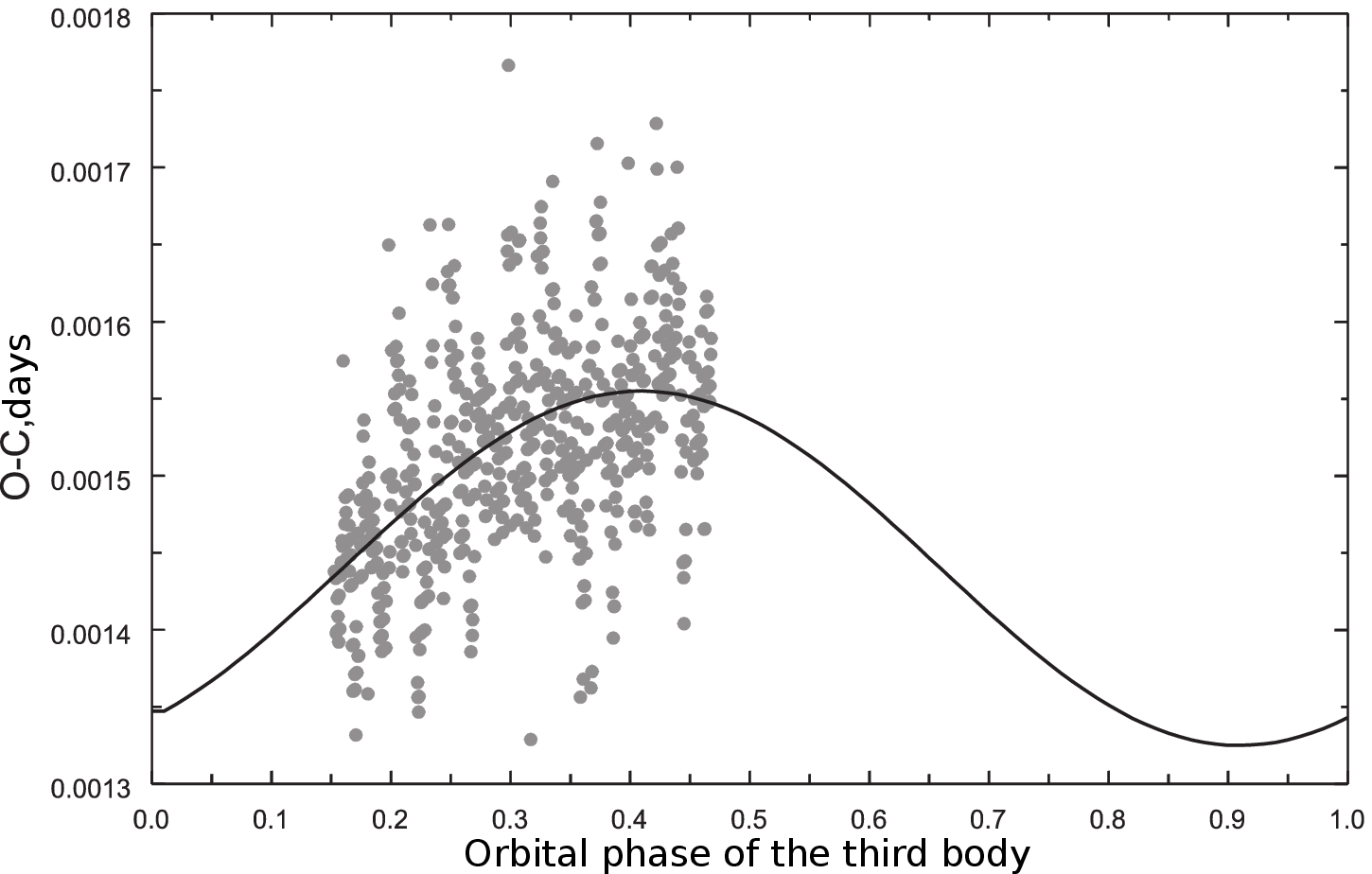}
\vspace{0cm}\caption{Same as Fig. \ref{p1} using the ephemeris \ref{ff3}.} \label{p2}
\end{figure}

\begin{figure}
\hspace{0cm} \epsfxsize=0.99\textwidth\centering \epsfbox{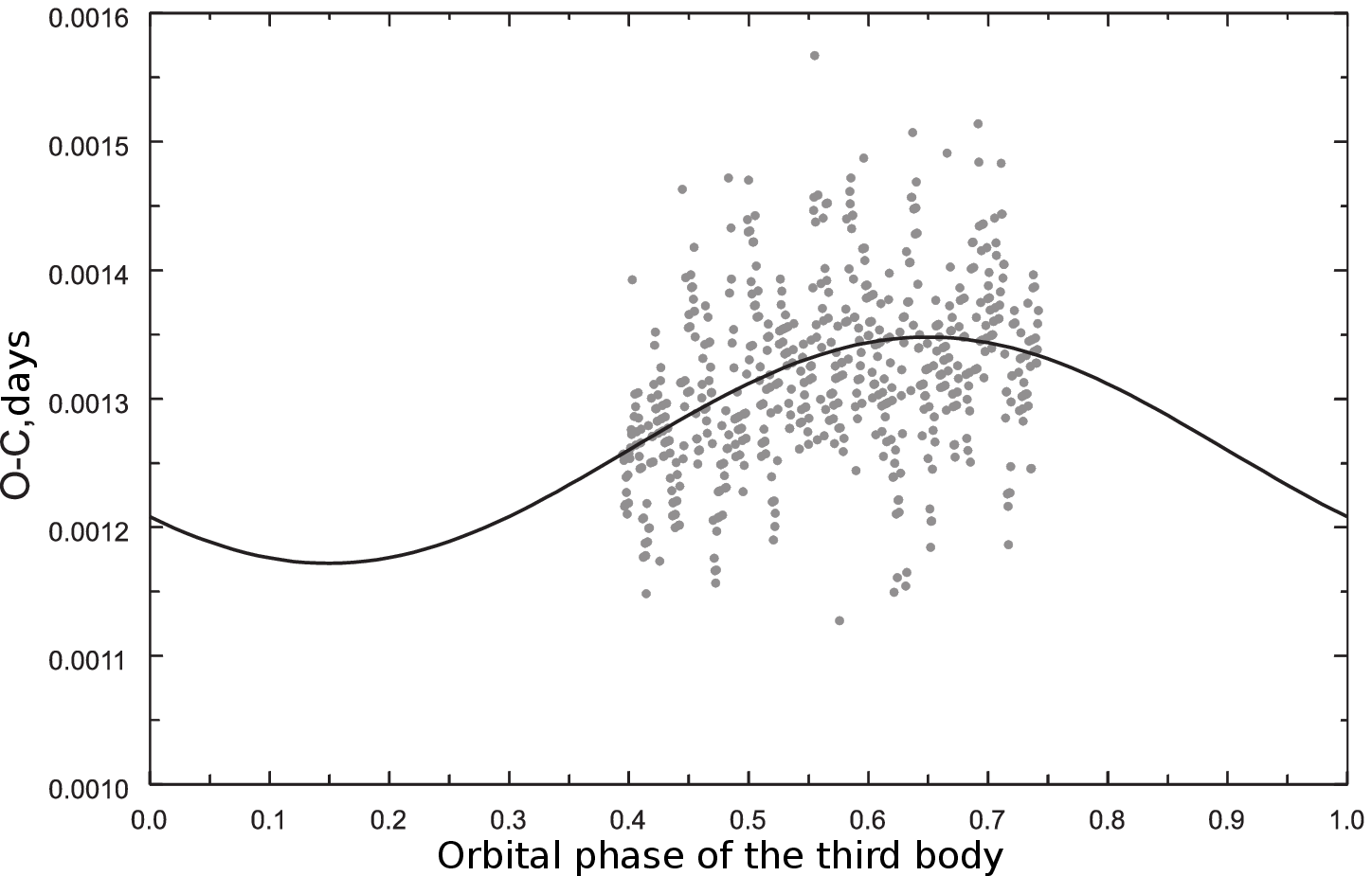}
\vspace{0cm}\caption{Same as Fig. \ref{p1} using the ephemeris \ref{ff4}.} \label{p3}
\end{figure}

We carried out our further analysis of the data obtained for the three periods
($P_{10}$,  $P_{11}$ , $P_{12}$). The large scatter of the O-C deviations limits our ability to obtain many parameters of the light-time effect. To minimize the number of parameters, we adopted
the simple hypothesis that the third body undergoes circular motion about the eclipsing binary. Using a
Fourier expansion\footnote{We applied the PERDET (PERiod DETermination) code \cite{breger1990}.}, we analyzed the O-C residuals obtained for each of the periods and the calculated parameters of the best-fit sine curve approximating the time dependence of the times of minima (the light-time effect). Table \ref{table4} presents the amplitudes and periods of this theoretical curve. Figure \ref{pow1} presents
the power spectrum for the O-C residuals calculated using the ephemeris \ref{eph-cris}, and Fig. \ref{pow2} displays a part of \ref{pow1} on a larger scale. The peak near a period of $\approx 2$
days corresponds to the orbital period of FL Lyr; this is clearly visible in Fig. \ref{skvazhnost}. A peak at about 5-6 yrs is also visible in Fig. \ref{pow1}; this is due to the light-time effect. It is difficult to search for a larger number of objects in the system due to the large scatter of the available times of minima, comparable to the amplitude of the light-time effect. The O-C calculations performed for all three periods ($P_{10}$, $P_{11}$ , $P_{12}$) demonstrate systematic deviations that can be explained as a light-time effect with a period somewhat larger than the entire time interval covered by the Kepler observations. Figures \ref{p1}, \ref{p2} and \ref{p3} show the O-C deviations of the times of minima as a function of the orbital phase of the third body.

If less than a half of the period of the light-time effect elapsed during the time covered by the Kepler
observations, an alternative explanation for the observed systematic shifts could be a variation of the
close-binary period ($dP \sim 10^{-5}$ days/year). The system has already been observed for a long time,
and period variations of this kind should already have been detected from the parabolic shape of the O-C
curve. During the time interval of the observations (almost 60 years), the FL Lyr period variations would
have already accumulated in the fourth place after the decimal point, and the period should be increasing,
while all the previously measured period values \cite{robinson1965}-\cite{samolyk2011} are not lower than those we have derived. This can be taken as evidence against the hypothesis that the system's period is varying, and that the times of minima exhibit variations due to the light-time effect.

\section{MASS OF THE THIRD BODY}

In the general case, accurately determining the mass of a body in a binary, and especially a multiple, system can require a dedicated, complex study. There is no sense in carrying out such estimates in
the framework of our current study, since the orbital period of the third body has not been accurately determined, and is longer than the time covered by the Kepler observations. Moreover, we were not able to
derive the orbital inclination of the third body relative to the orbital plane of the system. Therefore, we have obtained a simple lower limit for the third body's mass.

Since the orbital period of the third body is much longer than the orbital period of the central binary, we can use Kepler's third law to obtain a simple mass estimate\footnote{The orbital period of the third body is 7 years or more, compared to the 2-day orbital period of FL Lyr; stable orbits admitting application of Kepler's third law (for rough estimates, since there will definitely be perturbations of the third body's orbit) would begin with an orbital period of 20 days \cite{holman1999}.}

\begin{equation}
P_{orb}=0.1 \frac{a^{3/2}}{M^{1/2}}, \label{kepler}
\end{equation}

\noindent where $P_{orb}$ is the orbital period of the third body in days, $a$ the semi-major axis of the third body's orbit in solar radii, and $M$ the combined mass of the two components of FL Lyr and the third body in solar masses.

The sum of the component masses in the FL Lyr system is $\approx
2M_{\odot}$\cite{popper1986}. The orbital period of the third
body with the ephemeris \ref{eph-cris} is $\gtrsim 7$ years. According to \ref{kepler}, the semi-major axis of the orbit of the third body is $\gtrsim 1100 R_{\odot}$.The amplitude of the light-time
effect with the same ephemeris is 4.8 s. During this time, light traverses half the distance of the periodic shift of the FL Lyr binary due to the third body; i.e., the semi-major axis of the orbit of the FL Lyr system about the center of mass of the FL Lyr-third body system is approximately $2R_{\odot}$. Thus, the ratio of the third body's mass to the mass of the FL Lyr binary is $\approx 1/500$. We thus get the simple estimate for the third body's mass $2M_{\odot}/500\approx 4 M_J$. If the orbital period of the third body proves to be longer than our estimate, the estimated mass of this body will be lower; at the same time, the orbital inclination of the third body will increase its estimated mass. Note that the orbital planes for all eight known exoplanets in orbits around binaries are very close (within $1^{\circ}$) to the orbital planes of their parent binaries. Thus, our rough estimate of the planet's mass may prove to be close to the actual mass of this planet.

\section{CONCLUSIONS}

We have analyzed Kepler light curves for the eclipsing binary FL Lyr and detected the light-time
effect, indicating that the system probably contains a body with a mass of about four Jupiter masses, with
an orbital period around the close binary of $\ge 7$ yrs. Confirmation of this planet's existence will require long-term photometric observations of FL Lyr with an accuracy no worse than that of the Kepler data; otherwise, it will be necessary to analyze the radial-velocity curve of the system over a long time and with very high accuracy.

The times of minima we have derived from the light curve of FL Lyr (Table \ref{table3}) can be used in further studies of the system.

Discoveries of planets in close binary systems in recent years mean that the formation of two stellar
components during the collapse of a rotating protostellar cloud does not completely resolve the problem
of the inevitable angular-momentum excess in protostellar clouds. The formation of planets in circum-
stellar accretion-decretion disks remains necessary to completely resolve this problem. As a result, the
components of wide, as well as close, binary systems could have planets around them. This means that
most stars can possess planetary systems, so that the formation rate of planetary systems could be close to the star-formation rate (see, for instance, the recent paper \cite{ballard2014}). This rate for the Milky Way would thus be several planetary systems per year.

\section{ACKNOWLEDGEMENTS}

The authors thank A.I. Zakharov, S.E. Leontyev, and V.N. Sementsov, who developed software for the
computation of photometric elements of eclipsing binaries.

\clearpage

\begin{table}
\caption{Parameters derived from the Kepler light curves
and adopted in the calculations of the theoretical light
curve of FL Lyr  $r_1$, $r_2$ are the radii of the primary and
the secondary in units of the semi-major axis of FL Lyr; $e$ the orbital eccentricity; $\omega$ the
longitude of periastron; $L_1$, $L_2$ the luminosities of the
primary and secondary in units of the system's luminosity; $L_3$ the ``third light'' in units of the system's luminosity; $u_1$, $u_2$ limb-darkening coefficients for the primary and
secondary; and $\sigma$ the standard (O-C) deviation.
(O-C).}\label{table1}
\newcolumntype{C}{>{\centering\arraybackslash}m}
\begin{tabular}{|C{25mm}|C{30mm}|C{30mm}|}
\hline
Parameters & Value for & Values adopted \\
 & FL Lyr & in the computations \\
\hline
$r_1$ & 0.122-0.133 & 0.123 \\
$r_2$ & 0.118- 0.126 & 0.123 \\
$i$ & $85^{\circ}.3-86^{\circ}.5$ & $85^{\circ}.9$ \\
$e$ & 0-0.0002 & 0 \\
$\omega$ & 0-$360^{\circ}$ & 0 \\
$L_1$ & 0.520-0.610 & 0.596 \\
$L_2$ & 0.260-0.310 & 0.298 \\
$L_3$ & 0.008-0.22 & 0.106 \\
$u_1$ & 0.62 (fixed) & 0.62 (fixed) \\
$u_2$ & 0.68 (fixed) & 0.68 (fixed) \\
$\sigma_{O-C}$ & - & $0.^m00026$ \\
\hline
\end{tabular}
\end{table}

\begin{table}
\caption{Times of primary minima of FL Lyr from ground-
based observations, the O-C deviations were calculated
using the ephemeris
\ref{eph-cris}. }\label{table2}
\newcolumntype{C}{>{\centering\arraybackslash}m}
\begin{tabular}{|C{30mm}|C{30mm}|C{30mm}|}
\hline
Time of primary & & \\
minimum, HJD-2400000 & (O-C), days & Reference \\
\hline
38173.6400 &  0.00690 & \cite{robinson1965}  \\
38221.5525 &  0.00000 & \cite{cristaldi1965} \\
40038.1368 &  0.00350 & \cite{mallama1980} \\
40770.0011 &  0.00800 & \cite{mallama1980} \\
41133.7419 & -0.00303 & \cite{mallama1980} \\
41499.6666 & -0.00830 & \cite{mallama1980} \\
41865.6053 &  0.00050 & \cite{mallama1980} \\
42229.3473 & -0.00930 & \cite{mallama1980} \\
42595.2820 & -0.00454 & \cite{mallama1980} \\
42961.2145 & -0.00197 & \cite{mallama1980} \\
43690.8989 &  0.00070 & \cite{mallama1980} \\
44459.7830 & -0.00370 & \cite{scarfe1984} \\
45572.8218 & -0.00180 & \cite{scarfe1984} \\
46925.4551 & -0.00238 & \cite{keskin1989} \\
49909.5271 & -0.00191 & \cite{safar2000} \\
50654.4547 & -0.00312 & \cite{deeg2003} \\
51266.5174 & -0.00180 & \cite{agerer2000} \\
51440.7700 & -0.00155 & \cite{nelson2000} \\
52806.4725 & -0.00186 & \cite{agerer2003}\\
52911.0235 & -0.00227 & \cite{kim2006} \\
53209.4302 & -0.00273 & \cite{hubscher2005} \\
53531.7977 & -0.00208 & \cite{cook2005} \\
53555.7555 & -0.00400 & \cite{cook2005} \\
53612.3871 & -0.00439 & \cite{hubscher2006} \\
53673.3777 & -0.00211 & \cite{hubscher2006} \\
53684.2691 & -0.00149 & \cite{zejda2006} \\
54466.2261 & -0.00192 & \cite{hubscher2009} \\
54583.8446 & -0.00375 & \cite{samolyk2008} \\
54594.7376 & -0.00152 & \cite{samolyk2008} \\
54642.6572 & -0.00132 & \cite{samolyk2008} \\
55304.8135 & -0.00396 & \cite{samolyk2011} \\
55437.6827 & -0.00218 & \cite{samolyk2011} \\
\hline
\end{tabular}
\end{table}

\clearpage

\begin{longtable}{|c|c|}

\caption[Times of minima]{Times of primary minima of FL Lyr found from Kepler data, the O-C deviations were calculated using the
ephemeris \ref{eph-cris}. }\label{table3} \\

\hline \multicolumn{1}{|c|}{Moment, HJD-2400000} &
\multicolumn{1}{c|}{(O-C), days} \\ \hline
\endfirsthead

\multicolumn{2}{c}%
{{\bfseries \tablename\ \thetable{} -- continuation}} \\
\hline \multicolumn{1}{|c|}{Moment, HJD-2400000} &
\multicolumn{1}{c|}{(O-C), days} \\ \hline
\endhead

\hline \multicolumn{2}{|r|}{{continuation in the next page}} \\
\hline
\endfoot

\hline \hline
\endlastfoot

54965.02424 & -0.00113 \\
54967.20240 & -0.00113 \\
54969.38054 & -0.00114 \\
54971.55866 & -0.00118 \\
54973.73682 & -0.00117 \\
54975.91499 & -0.00115 \\
54978.09314 & -0.00116 \\
54980.27127 & -0.00118 \\
54982.44946 & -0.00115 \\
54984.62759 & -0.00117 \\
54986.80578 & -0.00114 \\
54988.98393 & -0.00114 \\
54991.16210 & -0.00113 \\
54993.34026 & -0.00112 \\
54995.51841 & -0.00112 \\
54997.69669 & -0.00100 \\
55004.23105 & -0.00110 \\
55006.40922 & -0.00109 \\
55008.58736 & -0.00110 \\
55010.76549 & -0.00113 \\
55012.94365 & -0.00112 \\
55017.29999 & -0.00109 \\
55019.47812 & -0.00111 \\
55021.65625 & -0.00114 \\
55023.83439 & -0.00115 \\
55026.01257 & -0.00113 \\
55028.19073 & -0.00112 \\
55030.36885 & -0.00115 \\
55032.54697 & -0.00119 \\
55034.72509 & -0.00122 \\
55036.90328 & -0.00119 \\
55039.08140 & -0.00122 \\
55041.25957 & -0.00121 \\
55043.43771 & -0.00122 \\
55045.61584 & -0.00125 \\
55047.79406 & -0.00118 \\
55049.97218 & -0.00121 \\
55052.15043 & -0.00112 \\
55054.32850 & -0.00120 \\
55056.50666 & -0.00120 \\
55060.86302 & -0.00115 \\
55063.04122 & -0.00110 \\
55065.21935 & -0.00113 \\
55067.39750 & -0.00113 \\
55069.57563 & -0.00115 \\
55071.75385 & -0.00109 \\
55073.93203 & -0.00106 \\
55076.11020 & -0.00105 \\
55078.28829 & -0.00111 \\
55080.46645 & -0.00111 \\
55082.64459 & -0.00112 \\
55084.82276 & -0.00110 \\
55087.00089 & -0.00113 \\
55091.35710 & -0.00123 \\
55093.53539 & -0.00109 \\
55095.71356 & -0.00108 \\
55097.89167 & -0.00112 \\
55100.06984 & -0.00111 \\
55102.24797 & -0.00113 \\
55104.42610 & -0.00115 \\
55106.60426 & -0.00115 \\
55108.78242 & -0.00114 \\
55110.96060 & -0.00112 \\
55113.13873 & -0.00114 \\
55115.31692 & -0.00111 \\
55117.49504 & -0.00114 \\
55119.67321 & -0.00113 \\
55121.85135 & -0.00114 \\
55124.02950 & -0.00114 \\
55126.20765 & -0.00115 \\
55128.38578 & -0.00117 \\
55132.74208 & -0.00118 \\
55134.92022 & -0.00120 \\
55137.09838 & -0.00119 \\
55139.27653 & -0.00119 \\
55141.45468 & -0.00120 \\
55143.63282 & -0.00121 \\
55145.81099 & -0.00120 \\
55147.98918 & -0.00116 \\
55150.16731 & -0.00119 \\
55152.34548 & -0.00117 \\
55156.70175 & -0.00121 \\
55158.87990 & -0.00121 \\
55161.05809 & -0.00118 \\
55163.23632 & -0.00110 \\
55165.41448 & -0.00110 \\
55167.59263 & -0.00110 \\
55169.77094 & -0.00095 \\
55171.94888 & -0.00116 \\
55174.12705 & -0.00115 \\
55176.30523 & -0.00112 \\
55178.48340 & -0.00110 \\
55180.66164 & -0.00102 \\
55187.19601 & -0.00111 \\
55189.37422 & -0.00106 \\
55191.55238 & -0.00105 \\
55193.73047 & -0.00111 \\
55195.90868 & -0.00106 \\
55198.08687 & -0.00102 \\
55200.26502 & -0.00103 \\
55202.44317 & -0.00103 \\
55204.62133 & -0.00103 \\
55206.79947 & -0.00104 \\
55208.97767 & -0.00100 \\
55211.15577 & -0.00105 \\
55213.33390 & -0.00107 \\
55215.51200 & -0.00113 \\
55217.69013 & -0.00115 \\
55219.86829 & -0.00115 \\
55222.04642 & -0.00117 \\
55224.22459 & -0.00116 \\
55228.58090 & -0.00116 \\
55235.11540 & -0.00112 \\
55237.29356 & -0.00111 \\
55239.47174 & -0.00109 \\
55243.82801 & -0.00113 \\
55246.00621 & -0.00108 \\
55248.18439 & -0.00105 \\
55250.36247 & -0.00113 \\
55252.54061 & -0.00114 \\
55254.71876 & -0.00115 \\
55256.89700 & -0.00106 \\
55259.07514 & -0.00108 \\
55261.25326 & -0.00111 \\
55263.43145 & -0.00108 \\
55265.60958 & -0.00110 \\
55267.78771 & -0.00112 \\
55269.96587 & -0.00112 \\
55272.14398 & -0.00116 \\
55274.32208 & -0.00122 \\
55278.67836 & -0.00125 \\
55280.85650 & -0.00126 \\
55283.03465 & -0.00127 \\
55285.21281 & -0.00126 \\
55287.39099 & -0.00123 \\
55289.56916 & -0.00122 \\
55291.74733 & -0.00120 \\
55293.92549 & -0.00120 \\
55296.10362 & -0.00122 \\
55298.28180 & -0.00120 \\
55300.45997 & -0.00118 \\
55302.63812 & -0.00118 \\
55304.81631 & -0.00115 \\
55306.99439 & -0.00122 \\
55311.35074 & -0.00118 \\
55313.52889 & -0.00119 \\
55315.70703 & -0.00120 \\
55317.88525 & -0.00114 \\
55320.06334 & -0.00120 \\
55322.24152 & -0.00117 \\
55324.41968 & -0.00117 \\
55326.59804 & -0.00096 \\
55328.77600 & -0.00116 \\
55330.95426 & -0.00105 \\
55335.31062 & -0.00100 \\
55337.48874 & -0.00104 \\
55339.66684 & -0.00109 \\
55344.02316 & -0.00108 \\
55346.20128 & -0.00111 \\
55348.37940 & -0.00115 \\
55350.55753 & -0.00117 \\
55352.73568 & -0.00118 \\
55354.91386 & -0.00115 \\
55357.09203 & -0.00113 \\
55359.27017 & -0.00115 \\
55361.44831 & -0.00116 \\
55363.62648 & -0.00115 \\
55365.80460 & -0.00118 \\
55367.98278 & -0.00116 \\
55372.33909 & -0.00116 \\
55374.51723 & -0.00117 \\
55376.69534 & -0.00121 \\
55378.87356 & -0.00115 \\
55381.05167 & -0.00119 \\
55383.22987 & -0.00115 \\
55385.40802 & -0.00115 \\
55387.58616 & -0.00117 \\
55389.76436 & -0.00112 \\
55391.94264 & -0.00100 \\
55394.12078 & -0.00101 \\
55396.29897 & -0.00097 \\
55398.47699 & -0.00111 \\
55400.65524 & -0.00101 \\
55402.83331 & -0.00110 \\
55405.01151 & -0.00105 \\
55407.18966 & -0.00106 \\
55409.36777 & -0.00110 \\
55411.54600 & -0.00102 \\
55413.72416 & -0.00102 \\
55415.90226 & -0.00107 \\
55418.08049 & -0.00100 \\
55420.25857 & -0.00107 \\
55422.43676 & -0.00104 \\
55424.61487 & -0.00108 \\
55428.97120 & -0.00106 \\
55431.14933 & -0.00108 \\
55433.32744 & -0.00113 \\
55435.50560 & -0.00112 \\
55437.68373 & -0.00115 \\
55439.86184 & -0.00119 \\
55442.04001 & -0.00118 \\
55444.21819 & -0.00115 \\
55446.39635 & -0.00115 \\
55448.57448 & -0.00117 \\
55450.75262 & -0.00118 \\
55452.93078 & -0.00118 \\
55455.10892 & -0.00119 \\
55457.28713 & -0.00114 \\
55459.46531 & -0.00111 \\
55461.64348 & -0.00110 \\
55463.82164 & -0.00109 \\
55465.99978 & -0.00110 \\
55468.17791 & -0.00113 \\
55470.35605 & -0.00114 \\
55472.53419 & -0.00116 \\
55474.71229 & -0.00121 \\
55476.89043 & -0.00123 \\
55479.06858 & -0.00123 \\
55481.24671 & -0.00126 \\
55483.42489 & -0.00123 \\
55485.60302 & -0.00125 \\
55487.78119 & -0.00124 \\
55489.95944 & -0.00114 \\
55494.31569 & -0.00120 \\
55496.49389 & -0.00116 \\
55498.67206 & -0.00114 \\
55500.85025 & -0.00111 \\
55503.02841 & -0.00110 \\
55505.20660 & -0.00106 \\
55507.38474 & -0.00108 \\
55509.56290 & -0.00107 \\
55511.74103 & -0.00110 \\
55513.91917 & -0.00111 \\
55516.09734 & -0.00110 \\
55518.27549 & -0.00110 \\
55520.45362 & -0.00112 \\
55522.63181 & -0.00109 \\
55524.80995 & -0.00110 \\
55526.98808 & -0.00113 \\
55529.16624 & -0.00112 \\
55531.34442 & -0.00110 \\
55533.52251 & -0.00116 \\
55535.70065 & -0.00118 \\
55537.87880 & -0.00118 \\
55540.05696 & -0.00117 \\
55542.23514 & -0.00115 \\
55546.59147 & -0.00113 \\
55548.76963 & -0.00112 \\
55550.94781 & -0.00110 \\
55570.55110 & -0.00120 \\
55572.72927 & -0.00118 \\
55574.90746 & -0.00114 \\
55577.08559 & -0.00117 \\
55579.26373 & -0.00118 \\
55581.44194 & -0.00113 \\
55583.62010 & -0.00112 \\
55585.79823 & -0.00115 \\
55587.97637 & -0.00116 \\
55590.15452 & -0.00117 \\
55592.33267 & -0.00117 \\
55598.86711 & -0.00119 \\
55601.04526 & -0.00120 \\
55603.22343 & -0.00118 \\
55605.40162 & -0.00115 \\
55607.57978 & -0.00114 \\
55609.75796 & -0.00112 \\
55611.93609 & -0.00114 \\
55614.11423 & -0.00115 \\
55616.29246 & -0.00108 \\
55618.47067 & -0.00102 \\
55620.64884 & -0.00101 \\
55622.82710 & -0.00090 \\
55625.00513 & -0.00103 \\
55627.18320 & -0.00111 \\
55629.36134 & -0.00112 \\
55631.53942 & -0.00120 \\
55633.71776 & -0.00101 \\
55642.43022 & -0.00117 \\
55644.60847 & -0.00108 \\
55646.78657 & -0.00113 \\
55648.96475 & -0.00110 \\
55651.14298 & -0.00103 \\
55653.32105 & -0.00111 \\
55655.49912 & -0.00120 \\
55657.67740 & -0.00107 \\
55659.85545 & -0.00118 \\
55662.03376 & -0.00102 \\
55664.21186 & -0.00108 \\
55666.39007 & -0.00102 \\
55668.56813 & -0.00111 \\
55670.74631 & -0.00109 \\
55672.92436 & -0.00119 \\
55677.28069 & -0.00117 \\
55679.45889 & -0.00113 \\
55681.63700 & -0.00117 \\
55683.81515 & -0.00117 \\
55685.99329 & -0.00119 \\
55688.17145 & -0.00118 \\
55690.34958 & -0.00121 \\
55692.52779 & -0.00115 \\
55694.70594 & -0.00116 \\
55696.88411 & -0.00114 \\
55699.06225 & -0.00116 \\
55701.24044 & -0.00112 \\
55703.41853 & -0.00118 \\
55705.59667 & -0.00120 \\
55707.77467 & -0.00135 \\
55709.95298 & -0.00120 \\
55712.13117 & -0.00116 \\
55714.30934 & -0.00115 \\
55716.48749 & -0.00115 \\
55718.66564 & -0.00116 \\
55720.84373 & -0.00122 \\
55723.02189 & -0.00121 \\
55727.37829 & -0.00112 \\
55729.55646 & -0.00111 \\
55731.73468 & -0.00104 \\
55733.91276 & -0.00112 \\
55738.26903 & -0.00115 \\
55740.44726 & -0.00108 \\
55742.62547 & -0.00102 \\
55744.80362 & -0.00103 \\
55746.98179 & -0.00101 \\
55749.15991 & -0.00105 \\
55751.33807 & -0.00104 \\
55753.51623 & -0.00104 \\
55755.69433 & -0.00109 \\
55757.87245 & -0.00112 \\
55760.05061 & -0.00112 \\
55762.22869 & -0.00119 \\
55764.40680 & -0.00124 \\
55766.58501 & -0.00118 \\
55768.76315 & -0.00120 \\
55770.94133 & -0.00117 \\
55773.11953 & -0.00113 \\
55775.29767 & -0.00114 \\
55777.47579 & -0.00117 \\
55779.65396 & -0.00116 \\
55781.83212 & -0.00115 \\
55784.01024 & -0.00119 \\
55786.18851 & -0.00107 \\
55788.36667 & -0.00107 \\
55790.54489 & -0.00100 \\
55792.72297 & -0.00107 \\
55794.90112 & -0.00108 \\
55797.07925 & -0.00110 \\
55799.25740 & -0.00111 \\
55801.43556 & -0.00110 \\
55803.61371 & -0.00111 \\
55805.79183 & -0.00114 \\
55807.96999 & -0.00114 \\
55810.14815 & -0.00113 \\
55812.32630 & -0.00113 \\
55814.50448 & -0.00111 \\
55816.68261 & -0.00113 \\
55818.86073 & -0.00117 \\
55821.03886 & -0.00119 \\
55823.21710 & -0.00111 \\
55825.39518 & -0.00118 \\
55827.57334 & -0.00118 \\
55829.75152 & -0.00115 \\
55831.92960 & -0.00122 \\
55836.28598 & -0.00115 \\
55838.46413 & -0.00116 \\
55840.64225 & -0.00119 \\
55842.82045 & -0.00115 \\
55844.99861 & -0.00114 \\
55847.17675 & -0.00115 \\
55849.35494 & -0.00112 \\
55851.53301 & -0.00120 \\
55853.71115 & -0.00122 \\
55855.88933 & -0.00119 \\
55858.06744 & -0.00124 \\
55860.24565 & -0.00118 \\
55862.42379 & -0.00120 \\
55864.60193 & -0.00121 \\
55866.78009 & -0.00120 \\
55868.95822 & -0.00123 \\
55871.13645 & -0.00115 \\
55873.31456 & -0.00120 \\
55875.49279 & -0.00112 \\
55877.67097 & -0.00110 \\
55879.84907 & -0.00115 \\
55882.02721 & -0.00117 \\
55884.20530 & -0.00123 \\
55886.38349 & -0.00119 \\
55888.56164 & -0.00120 \\
55890.73973 & -0.00126 \\
55892.91789 & -0.00126 \\
55895.09595 & -0.00135 \\
55897.27421 & -0.00125 \\
55899.45237 & -0.00124 \\
55901.63047 & -0.00129 \\
55905.98673 & -0.00134 \\
55908.16495 & -0.00128 \\
55910.34310 & -0.00128 \\
55912.52125 & -0.00129 \\
55914.69954 & -0.00115 \\
55916.87759 & -0.00126 \\
55919.05580 & -0.00120 \\
55921.23397 & -0.00118 \\
55923.41208 & -0.00123 \\
55925.59032 & -0.00114 \\
55927.76846 & -0.00116 \\
55929.94663 & -0.00114 \\
55934.30273 & -0.00135 \\
55936.48115 & -0.00109 \\
55938.65905 & -0.00134 \\
55940.83741 & -0.00113 \\
55943.01557 & -0.00113 \\
55945.19372 & -0.00113 \\
55947.37191 & -0.00110 \\
55949.55006 & -0.00110 \\
55951.72812 & -0.00120 \\
55953.90642 & -0.00105 \\
55956.08457 & -0.00105 \\
55958.26278 & -0.00100 \\
55960.44078 & -0.00115 \\
55962.61903 & -0.00106 \\
55964.79718 & -0.00106 \\
55966.97532 & -0.00108 \\
55969.15349 & -0.00106 \\
55971.33167 & -0.00104 \\
55973.50978 & -0.00108 \\
55975.68789 & -0.00112 \\
55977.86600 & -0.00117 \\
55980.04415 & -0.00117 \\
55982.22232 & -0.00116 \\
55984.40043 & -0.00120 \\
55990.93486 & -0.00124 \\
55993.11305 & -0.00120 \\
55995.29120 & -0.00120 \\
55997.46934 & -0.00122 \\
55999.64750 & -0.00121 \\
56001.82566 & -0.00121 \\
56004.00383 & -0.00119 \\
56006.18199 & -0.00119 \\
56008.36016 & -0.00117 \\
56010.53822 & -0.00126 \\
56012.71645 & -0.00119 \\
56014.89457 & -0.00122 \\
56017.07265 & -0.00130 \\
56019.25077 & -0.00133 \\
56021.42895 & -0.00131 \\
56023.60710 & -0.00131 \\
56025.78530 & -0.00127 \\
56027.96348 & -0.00124 \\
56030.14168 & -0.00119 \\
56032.31980 & -0.00123 \\
56034.49793 & -0.00125 \\
56036.67620 & -0.00114 \\
56038.85433 & -0.00116 \\
56041.03247 & -0.00118 \\
56043.21064 & -0.00116 \\
56045.38880 & -0.00116 \\
56047.56695 & -0.00116 \\
56049.74509 & -0.00117 \\
56051.92322 & -0.00120 \\
56054.10136 & -0.00121 \\
56056.27953 & -0.00120 \\
56058.45768 & -0.00120 \\
56060.63583 & -0.00121 \\
56062.81401 & -0.00118 \\
56064.99215 & -0.00119 \\
56067.17031 & -0.00119 \\
56069.34844 & -0.00121 \\
56071.52663 & -0.00118 \\
56073.70473 & -0.00123 \\
56075.88309 & -0.00103 \\
56080.23923 & -0.00120 \\
56082.41739 & -0.00119 \\
56084.59558 & -0.00115 \\
56086.77377 & -0.00112 \\
56088.95187 & -0.00117 \\
56091.13003 & -0.00117 \\
56093.30819 & -0.00116 \\
56095.48629 & -0.00122 \\
56097.66443 & -0.00123 \\
56099.84256 & -0.00126 \\
56102.02071 & -0.00126 \\
56104.19885 & -0.00127 \\
56108.55520 & -0.00123 \\
56110.73337 & -0.00122 \\
56112.91154 & -0.00120 \\
56115.08969 & -0.00121 \\
56117.26788 & -0.00117 \\
56119.44606 & -0.00114 \\
56121.62421 & -0.00115 \\
56130.33683 & -0.00115 \\
56132.51495 & -0.00118 \\
56134.69314 & -0.00115 \\
56136.87126 & -0.00118 \\
56141.22754 & -0.00121 \\
56143.40564 & -0.00126 \\
56145.58383 & -0.00123 \\
56147.76194 & -0.00127 \\
56149.94015 & -0.00122 \\
56152.11831 & -0.00121 \\
56154.29644 & -0.00124 \\
56156.47455 & -0.00128 \\
56158.65285 & -0.00113 \\
56160.83101 & -0.00113 \\
56163.00918 & -0.00111 \\
56165.18734 & -0.00111 \\
56167.36547 & -0.00113 \\
56171.72170 & -0.00121 \\
56176.07805 & -0.00117 \\
56178.25616 & -0.00121 \\
56182.61266 & -0.00102 \\
56184.79079 & -0.00105 \\
56186.96889 & -0.00110 \\
56189.14696 & -0.00119 \\
56191.32514 & -0.00116 \\
56193.50333 & -0.00112 \\
56195.68142 & -0.00119 \\
56197.85966 & -0.00110 \\
56200.03782 & -0.00110 \\
56202.21585 & -0.00122 \\
56206.57218 & -0.00120 \\
56208.75036 & -0.00118 \\
56210.92850 & -0.00119 \\
56213.10672 & -0.00112 \\
56215.28484 & -0.00116 \\
56217.46300 & -0.00115 \\
56219.64117 & -0.00114 \\
56221.81930 & -0.00116 \\
56223.99745 & -0.00117 \\
56226.17558 & -0.00119 \\
56228.35372 & -0.00120 \\
56230.53189 & -0.00119 \\
56232.71004 & -0.00119 \\
56234.88821 & -0.00118 \\
56237.06636 & -0.00118 \\
56239.24460 & -0.00110 \\
56241.42268 & -0.00117 \\
56243.60089 & -0.00112 \\
56245.77903 & -0.00113 \\
56252.31344 & -0.00118 \\
56254.49160 & -0.00118 \\
56256.66976 & -0.00117 \\
56258.84793 & -0.00116 \\
56261.02618 & -0.00106 \\
56263.20430 & -0.00110 \\
56265.38245 & -0.00110 \\
56267.56055 & -0.00115 \\
56269.73872 & -0.00114 \\
56271.91687 & -0.00114 \\
56274.09496 & -0.00121 \\
56276.27306 & -0.00126 \\
56278.45124 & -0.00124 \\
56280.62939 & -0.00124 \\
56282.80746 & -0.00132 \\
56284.98561 & -0.00133 \\
56287.16373 & -0.00136 \\
56289.34193 & -0.00132 \\
56291.52008 & -0.00132 \\
56293.69826 & -0.00130 \\
56295.87646 & -0.00125 \\
56298.05464 & -0.00123 \\
56300.23279 & -0.00123 \\
56302.41098 & -0.00119 \\
56306.76730 & -0.00118 \\
56308.94545 & -0.00119 \\
56322.01433 & -0.00123 \\
56324.19246 & -0.00126 \\
56326.37065 & -0.00122 \\
56328.54883 & -0.00120 \\
56330.72696 & -0.00122 \\
56332.90509 & -0.00125 \\
56335.08323 & -0.00126 \\
56337.26137 & -0.00127 \\
56339.43956 & -0.00124 \\
56341.61770 & -0.00125 \\
56343.79586 & -0.00125 \\
56345.97401 & -0.00125 \\
56348.15220 & -0.00122 \\
56350.33032 & -0.00125 \\
56352.50855 & -0.00118 \\
56354.68662 & -0.00126 \\
56356.86482 & -0.00121 \\
56361.22111 & -0.00123 \\
56363.39919 & -0.00131 \\
56365.57734 & -0.00131 \\
56367.75560 & -0.00121 \\
56369.93379 & -0.00117 \\
56372.11196 & -0.00116 \\
56374.29005 & -0.00122 \\
56376.46825 & -0.00117 \\
56378.64637 & -0.00121 \\
56380.82450 & -0.00123 \\
56383.00266 & -0.00123 \\
56385.18082 & -0.00122 \\
56387.35900 & -0.00120 \\
56389.53716 & -0.00119 \\
\hline
\end{longtable}

\clearpage

\begin{table}
\caption{Amplitude of the theoretical curve and the orbital
period of the third body obtained from a Fourier expansion
using three values of the orbital period of FL Lyr}\label{table4}
\newcolumntype{C}{>{\centering\arraybackslash}m}
\begin{tabular}{|C{25mm}|C{30mm}|C{30mm}|}
\hline
Orbital period of FL Lyr, days & Light-time effect amplitude, s & Light-time effect period, years \\
\hline
2.17815440 & 4.8 & 7.2 \\
2.17815408 & 9.9 & 12.4 \\
2.17815414 & 7.6 & 11.3 \\
\hline
\end{tabular}
\end{table}

\clearpage

\end{document}